\documentclass[journal,10pt]{IEEEtran}

\usepackage{psfrag}
\usepackage[utf8]{inputenc} 
\usepackage[T1]{fontenc}    
\usepackage{hyperref}       
\usepackage{url}            
\usepackage{booktabs}       
\usepackage{amsfonts}       
\usepackage{nicefrac}       
\usepackage{microtype}      

\usepackage[final]{graphicx}
\graphicspath{{Images//}} 

\usepackage{psfrag}
\usepackage{epstopdf}
\usepackage{amsfonts}
\usepackage{mathrsfs}
\usepackage{pifont}
\usepackage{verbatim}
\usepackage{upgreek}
\usepackage{soul,color}
\usepackage[usenames,dvipsnames]{xcolor}
\usepackage{caption}
\usepackage{algorithmic}
\usepackage{algorithm}
\usepackage{mdwmath}
\usepackage{mdwtab}
\usepackage{amssymb,amsmath}
\usepackage{amsmath}
\usepackage{amsthm}
\usepackage{epstopdf}
\usepackage{tikz}
\usepackage{scalefnt}
\usepackage{subfigure}
\usepackage{MnSymbol}
\usepackage{setspace}
\usepackage{mathtools}
\usepackage{psfrag}
\usepackage{cite}

\label{newcommand}
\newcommand{\bzero}{\mbox{\boldmath{$0$}}}

\label{English Chars}
\newcommand{\bA}{\textbf{A}}
\newcommand{\ba}{\textbf{a}}

\newcommand{\be}{\textbf{e}}
\newcommand{\bF}{\textbf{F}}

\newcommand{\bI}{\textbf{I}}

\newcommand{\bR}{\textbf{R}}

\newcommand{\bu}{\textbf{u}}
\newcommand{\bV}{\textbf{V}}
\newcommand{\bv}{\textbf{v}}

\newcommand{\bX}{\textbf{X}}
\newcommand{\bx}{\textbf{x}}
\newcommand{\bY}{\textbf{Y}}
\newcommand{\by}{\textbf{y}}
\newcommand{\bZ}{\textbf{Z}}

\label{Roman Chars}

\newcommand{\btheta}{\mbox{\boldmath{$\theta$}}}

\newcommand{\bPhi}{\mbox{\boldmath{$\Phi$}}}
\newcommand{\bphi}{\mbox{\boldmath{$\phi$}}}

\newcommand{\bPi}{\mbox{\boldmath{$\Pi$}}}

\def\So{\mathcal{S}_{\rm OPT}}
\def\Si{\mathcal{S}_i}
\def\Sc{\mathcal{S}}

\label{Theorems}
\newtheorem{theorem}{Theorem}

\newtheorem{corollary}{Corollary}

\newtheorem{lemma}{Lemma}
\newtheorem{proposition}{Proposition}
\newtheorem{definition}{Definition}

\theoremstyle{remark}

\begin{document}

    \title{Revisiting Matching Pursuit:\\ Beyond Approximate Submodularity}
    
	\author{%
	\IEEEauthorblockN{Ehsan Tohidi\IEEEauthorrefmark{1}\IEEEauthorrefmark{2},
		Mario Coutino\IEEEauthorrefmark{3},
		David Gesbert\IEEEauthorrefmark{4}}\\
	\IEEEauthorblockA{\IEEEauthorrefmark{1}%
		Fraunhofer HHI, Berlin, Germany}\\ 
\IEEEauthorblockA{\IEEEauthorrefmark{2} Technische Universität Berlin, Berlin, Germany}\\
\IEEEauthorblockA{\IEEEauthorrefmark{3}%
		Faculty of Electrical Engineering, Mathematics and Computer Science, Delft University of Technology, Delft, The Netherlands 
		}\\
	\IEEEauthorblockA{\IEEEauthorrefmark{4}%
		Communication Systems Department, EURECOM,
		Biot, France}
	
}

	\maketitle

\begin{abstract}
	We study the problem of selecting a subset of vectors from a large set, to obtain the best signal representation over a family of functions. Although greedy methods have been widely used for tackling this problem and many of those have been analyzed under the lens of (weak) submodularity, none of these algorithms are explicitly devised using such a functional property. Here, we revisit the vector-selection problem and introduce a function which is shown to be submodular in expectation. This function does not only guarantee near-optimality through a greedy algorithm in expectation, but also alleviates the existing deficiencies in commonly used matching pursuit (MP) algorithms. We further show the relation between the single-point-estimate version of the proposed greedy algorithm and MP variants. Our theoretical results are supported by numerical experiments for the angle of arrival estimation problem, a typical signal representation task; the experiments demonstrate the benefits of the proposed method with respect to the traditional MP algorithms.
	
\end{abstract}

\section{Introduction}
In this paper, we are interested in revisiting the ubiquitous problem of signal representation: from a given set of vectors, the problem consists in selecting a subset of $K$ of them which \emph{best} represent another vector of interest, i.e., the target signal. This problem has found many applications in signal processing and machine learning~\cite{258082}. Within the realm of signal processing, the vectors could be delayed-versions of a reference signal, and the goal is to represent another with few of them; This problem is found for instance in direction of arrival estimation (DOA)~\cite{van2004optimum}. In machine learning, the vectors often relate to features, and we want to make a prediction of a given phenomenon using only a small part of the available features. Typically, the involved vectors are referred to as \emph{atoms} and the set of atoms is termed \emph{dictionary}. 

When dictionaries are composed of orthonormal atoms, the selection problem naturally reduces to representing the target signal as the superposition of basis vectors; see, e.g.,~\cite{akansu1991signal}. However, in the case of redundant dictionaries, i.e., non-orthonormal atoms, the uniqueness of the representation is not necessarily guaranteed. Due to interpretability motivations or constraints given by a priori information, a parsimonious representation of the target signal is usually preferred, which means that the representation with the least amount of atoms is searched. Although finding the global optimal solution to this problem is computationally intractable, greedy algorithms, e.g., matching pursuit (MP) and orthogonal MP (OMP), efficiently find such a solution under appropriate conditions~\cite{tropp2007signal}.

Aligned with the goal of the signal representation problem, MP-based methods aim to compute a linear expansion of the target signal in terms of a few atoms of the given dictionary. While MP constructs its solution by successive approximations of the target signal with orthogonal projections on elements of the dictionary~\cite{258082}, OMP builds its solution by ensuring that the residual is orthogonal to the span of previously added atoms~\cite{342465}. Despite their differences, both methods normalize all the atoms in the beginning, and, at each step, \emph{greedily} select the next atom, among the unselected ones, with the maximum inner product with the residual. Though these methods exhibit good performance in several instances~\cite{tropp2007signal}, the fact that a part of each unselected atom could lie over the span of the currently selected atoms affects negatively the atom selection step because the remaining parts are not normalized anymore.

Although there have been many extensions of MP methods, e.g.,~\cite{1001652,1324706,Needell2009,6302206}, none of these extensions provide guarantees on the optimality of their solution for general problems. In addition, their performance guarantees (if any) are only applicable to special cases, e.g., signals with sparse representation or dictionaries with specific characteristics \cite{258082,342465,1001652,1324706,Needell2009,6302206,5550495}. 

Another family of methods, intended for the same goal, determine the representation using some relaxations/modifications of the original problem~\cite{chen2001atomic,471413,6719509,5361489,8683203}. For instance, the method of frames (MOF), among all solutions, picks out one whose representation coefficients have the minimum $\ell_2$-norm, while the principle of basis pursuit (BP) is to find a representation of the signal whose coefficients have the minimum $\ell_1$-norm \cite{chen2001atomic,471413}. Yet, no guarantee is presented or the guarantees are valid only for special cases.

Traditional ways to develop theoretical performance bounds for subset selection algorithms mostly involve the spectral conditions of matrices related to the input data, see, e.g.,~\cite{tropp2004greed}. However, the functional property of submodularity can also be leveraged to analyze and produce performance guarantees. These efforts have led to multiplicative approximation bounds for methods such as forward regression (FR) and OMP improving past known bounds~\cite{das2011submodular}.

Submodularity is a characteristic of set functions that exhibit the so-called \emph{diminishing returns} property \cite{tohidi2020submodularity}. Employing greedy algorithms, the maximization of a submodular function subject to cardinality and matroid constraints (which are imposed in many practical signal representation problems) can be performed near-optimally within a $1-1/\rm{e}$ factor \cite{nemhauser1978analysis,calinescu2007maximizing,calinescu2011maximizing}. Due to the near-optimality guarantees and the conceptually simple algorithms, submodular optimization has appeared in a wide variety of applications; e.g., antenna selection in multi-input multi-output systems \cite{8537943}, virtual controller placement in 5G software defined networks \cite{tohidi2021near}, resource allocation in wireless communication \cite{7208844}, sparse sampler design for detection \cite{8379441}, intelligent reconfigurable surfaces placement \cite|{tohidi2023near}, distributed machine learning \cite{mirzasoleiman16a}, regression \cite{NIPS2018_7286}, online approximation algorithms \cite{jegelka2011online}, multi-objective optimization \cite{NIPS2018_8159}, and active learning \cite{wei2015submodularity}. In addition, \cite{Krause3104395} has proposed dictionary selection algorithms exploiting approximate submodularity where the performance bounds are dependent on the dictionary.

Although submodularity has been used to explain the good performance of greedy algorithms for the signal representation problem, the analyzed algorithms themselves have not in fact been devised explicitly using submodularity as a foundation. In this paper, we revisit the problem of signal representation from a new direction, and introduce a submodular objective function to devise a greedy algorithm, referred as submodular matching pursuit (SMP), that $(i)$ has provable near-optimality guarantees and $(ii)$ overcomes the shortcomings of the OMP algorithm. Using submodularity as a stepping stone, we discuss the possibility to naturally extend the signal representation problem to instances with constraints, namely knapsack and matroid constraints while still ensuring near-optimality performance. Although other structured constraints have been considered in the past, e.g., block sparsity, to the best of our knowledge, this is the first time that the aforementioned constraints are considered in the signal representation setting. Our main contributions can be summarized as follows:
\begin{enumerate}
    \item We revisit the problem of signal representation and motivated from a wide range of applications, we introduce a submodular objective set function for this problem.
    \item We propose the SMP algorithm; a greedy algorithm with a $(1-\rm{e}^{-1})$-approximation ratio for the signal representation problem, in expectation.
    \item We show that the so-called optimized orthogonal pursuit (OOMP) algorithm~\cite{1001652} can be interpreted as the single-point-estimate version of SMP.
    \item We show the near-optimality (in probability) when using the finite-sample version of the introduced submodular function. This result hence establishes the near-optimality of OOMP.
    
    \item We discuss the extension of the signal representation problem to instances with knapsack and matroid constraints and argue that slightly modified SMP algorithms return solutions with $\frac{1}{2}(1-\rm{e}^{-1})$- and $1/2$-approximation ratios, respectively.
\end{enumerate}

\textbf{Notation.} Boldface lower(upper)case denote column vectors, e.g., $\ba$ (matrices $\bA$). Calligraphic letters are used for sets and $|\cal{A}|$ denotes the cardinality of set $\cal{A}$; $\setminus$ is used as the operator of set subtraction. The Hermitian operator is denoted by $(\cdot)^H$. Given a dictionary matrix $\bPhi$, and a subset $\mathcal{S}$ of indices of columns of $\bPhi$, $\bPhi_{\mathcal{S}}$ denotes a matrix composed of the columns of $\bPhi$ indexed by $\mathcal{S}$. $\operatorname{span}\{\bA\}$ stands for the span of columns of $\bA$; and $\bPi_\mathcal{S} = \bPhi_{\mathcal{S}}(\bPhi_{\mathcal{S}}^H\bPhi_{\mathcal{S}})^{-1}\bPhi_{\mathcal{S}}^H$ is the projection matrix onto $\operatorname{span}\{\bPhi_{\mathcal{S}}\}$. $\by_{\mathcal{S}}$ is the projection of $\by$ over $\operatorname{span}\{\bPhi_{\mathcal{S}}\}$. $<\cdot,\cdot>$ and $||\cdot||_2$ denote the inner product and the $\ell_2$-norm, respectively.

\section{Preliminaries}

Consider a set of $N$ $M$-dimensional \emph{atoms}, $\{\bphi_i\in\mathbb{C}^M\}_{i=1}^N$, being represented by a matrix $\bPhi\in\mathbb{C}^{M\times N}$. This matrix is usually referred as \emph{dictionary}. In the signal representation problem, we aim to find an accurate representation of a \emph{target signal} $\by\in\mathbb{C}^{M\times 1}$ over $\operatorname{span}\{\bPhi\}$ such that only a few atoms are involved. In other words, we look for a representation $\by_{\cal{S}}$ with $|\cal{S}|$ as small as possible. 

A related problem, but with a different context, is the so-called signal recovery problem. Given a measurement vector $\by$, the goal is to find the atoms and corresponding coefficients that best reconstruct the signal. The recovery problem is modeled as follows
\begin{equation}
\by = \bPhi \bx + \be,
\label{equ:recovery}
\end{equation}
where $\by\in\mathbb{C}^{M\times 1}$, $\bPhi\in\mathbb{C}^{M\times N}$, $\bx\in \mathbb{C}^{N\times 1}$, and $\be\in \mathbb{C}^{M\times 1}$ are the measurement vector, dictionary matrix, unknown coefficients vector, and noise vector, respectively. This problem reduces to estimate the non-zero coefficients of $\bx$, given the measurement vector $\by$, under the assumption that $\bPhi$ is known.

As both signal representation and recovery problems have the same goal -selecting a subset of atoms with the minimum cardinality that best represents the signal- though in different contexts, our problem formulation and the proposed algorithms in this work apply to both of them.

\subsection{Submodularity}

Here, we list a series of definitions needed in this paper and a set of theoretical results, within submodular analysis, which we use to sustain the claims of this work.

First, we introduce the formal definition of submodularity.

\begin{definition}
	(Submodularity) A set function $f: 2^{\mathcal{N}}\to \mathbb{R}$ defined over the ground set $\mathcal{N}=\{1,2,\cdots,N\}$ is submodular if for every $\mathcal{S}\subseteq\mathcal{N}$ and for all $a,b\in\mathcal{N}\setminus\mathcal{S}$ it holds that
	\begin{equation}
	f(\mathcal{S}\cup\{a\}) - f(\mathcal{S}) \geqslant f(\mathcal{S}\cup\{a,b\}) - f(\mathcal{S}\cup\{b\}).
	\label{equ:subcondorig}
	\end{equation}	
\end{definition}

A common property of set functions appearing in the signal representation problem is monotonicity.

\begin{definition}
	(Monotonicity) A set function $f: 2^{\mathcal{N}}\to \mathbb{R}$ defined over the ground set $\mathcal{N}=\{1,2,\cdots,N\}$ is monotonic if for every $\mathcal{A}\subseteq\mathcal{B}\subseteq\mathcal{N}$, $f(\mathcal{A}) \leqslant f(\mathcal{B})$.
\end{definition}

Finally, a set function is said to be \emph{normalized} if $f(\emptyset)=0$.

We now formally define the \emph{constraint subset selection problem}.

\begin{definition} (Constrained subset selection) Given a ground set $\mathcal{N}$, a set function $f$, and a set of sets $\mathcal{I}$, find a set satisfying $\mathcal{S}\subseteq\mathcal{N}$ and $\mathcal{S}\in \mathcal{I}$, maximizing $f(\mathcal{S})$.
\end{definition}

From the above definition, it is clear that the signal representation problem (described before) is an instance of the constrained subset selection problem, where the function $f$ measures how well the set $\mathcal{S}$ of atoms represents the target signal. The set of sets $\mathcal{I}$, in the standard version of the problem, are all subsets of $\mathcal{N}$ whose cardinality is smaller than a prescribed $K$, i.e., $\mathcal{I}_{K} :=\{\mathcal{A}:\mathcal{A}\subseteq\mathcal{N},\,|\mathcal{A}|\leq K\}$.

{In order to illustrate constrained subset selection problems in Definition 3 more clearly, here, we briefly discuss the so-called sensing coverage problem. This example application will be used throughout the paper to make concrete constraints discussed later in the paper.}

\begin{figure}
	\centering
	\includegraphics[width=.5\textwidth]{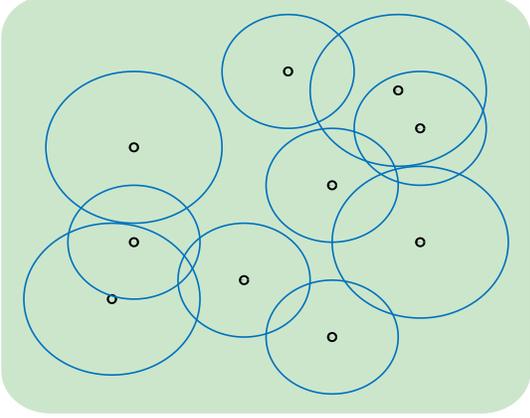}		
	\caption{Illustration of the sensing coverage problem with the coverage of each sensor. The goal is to select a subset of sensors to maximize the covered area.}
	\label{warmup_example}
\end{figure}

{Consider that there exists a set of possible locations of sensors (i.e., the ground set) with functioning areas that determine the area around each sensor from where useful information can be measured. 
Since these sensors are costly, we desire to optimize the deployment locations in order to maximize their effectiveness while keeping the cost at bay. Therefore, the optimization problem results in a coverage problem where the objective function is to maximize the covered area of the sensing field subject to a limited number of selected sensors. An example of such a sensor selection problem is presented in Fig. \ref{warmup_example}. This setting, naturally, leads to a \emph{constrained subset selection} \cite{tohidi2020submodularity}.}

{Though at first glance, the sensing coverage problem seems to be disconnected from the signal representation problem, it is possible to establish a direct relation between these two problems as follows. Consider the correspondence between the sensors to the columns of the dictionary,
and the target signal to the target field (i.e., the rectangular field in this case). Clearly, some atoms/sensors are correlated/have coverage overlap and some are uncorrelated/separated. We aim to represent/fill the target signal/field with the atoms/sensors.}

Now that the motivation behind the discussed setting is in order, we now provide a well-known result with respect to the solution of this kind of problems using a greedy heuristic. That is, when the set function $f$ is submodular, and $\mathcal{I}$ takes a particular form, we have the following known result.

\begin{theorem}\label{the1}\cite{nemhauser1978analysis}
	Let $f: 2^{\mathcal{N}}\to \mathbb{R}$ be a normalized, monotonic, submodular set function defined on the subsets of a finite ground set $\mathcal{N}$. Let $\mathcal{S}_{{\rm G}}$ be a set of $K$ elements selected by Algorithm \ref{GreedyAlgo1}. Then,
	\begin{equation}
	f(\mathcal{S}_{{\rm G}}) \geqslant (1-{\rm{e}}^{-1}) f(\mathcal{S}_{{\rm OPT}}),
	\end{equation}
	where $\mathcal{S}_{{\rm OPT}}$ is the solution to the cardinality-constrained subset selection problem with $\mathcal{I}=\mathcal{I}_{K}$.
\end{theorem}

\subsection{Matching Pursuit Algorithms}

We briefly review the two main variants of MP methods, namely MP and OMP, to put them in context with respect to Algorithm 1.

\textbf{Matching Pursuit.} The original version of MP iteratively constructs a solution by substituting line $2$ in Algorithm~\ref{GreedyAlgo1} by 
\begin{align}\label{eq.mprule}
    i^* = i_k := \underset{i\not\in\mathcal{S}_{k-1}}{\arg\max}\; \vert\langle R_k,\bphi_i\rangle\vert,
\end{align}
where $R_{k+1} := R_{k} - \alpha_k\bphi_{i_k}$, with $\alpha_k:= \vert\langle R_k,\bphi_{i_k}\rangle\vert$; and $R_1 := \by$. 

\textbf{Orthogonal Matching Pursuit.} In this refined version of MP, the selection rule is the same as that of~\eqref{eq.mprule}, but with a slightly different definition for the $k$th residual, i.e.,
\begin{align}\label{eq.omprule}
    R_k := \by - \by_{\mathcal{S}_{k-1}}.
\end{align}

Although OMP improves on MP, in general, the $k$th-selected atom cannot be guaranteed to be completely outside of $\operatorname{span}\{\bPhi_{\mathcal{S}_{k-1}}\}$. This characteristic hinders the ability of OMP (and subsequently MP) to find the best representation for the target signal in many instances. 


\subsection{Pitfalls of OMP: A Motivating Toy Example}
To illustrate the aforementioned problem of OMP, let us study the following toy problem. Consider $M=N=3$ and $\by=[1000,10,1]^H$. Let the dictionary, with normalized columns, be given as
\begin{equation}
\bPhi=
\begin{bmatrix}
1 & 0.9959 & 0\\ 0 & 0.09 & 0 \\ 0 & 0 & 1
\end{bmatrix}.
\label{equ:toy1}
\end{equation}
Assume a cardinality constraint which limits the representation to maximum $K~=2$ atoms. In the first iteration, OMP selects the first column as it has the maximum inner product with the signal $\by$. Removing the projected part, i.e., $\by_{\{1\}}$, the residual is equal to $R_2=[0,10,1]^T$. A quick calculation of the inner product between $R_2$ with the two remaining columns gives
\begin{equation}
\begin{aligned}
\langle R_2,\bphi_2\rangle &= 0.9, \\
\langle R_2,\bphi_3\rangle &= 1.
\end{aligned}
\end{equation}
Therefore, at the second iteration OMP selects the third column forming the solution $\mathcal{S}_{{\rm OMP}}=\{1,3\}$. This leaves a final residual of $R_3 = [0,10,0]^H$ with norm $10$, while, if we select the first two columns, i.e., $\mathcal{S}'=\{1,2\}$, the final residual would have a norm of $1$, i.e., $R_3'=[0,0,1]^H$.

The reason for this deficiency in the selection of vectors is that at the second iteration, the second column has a non-zero inner product with the first column which is already selected. This causes the normalization of columns in the initialization phase unfair.

In the following, using~\eqref{eq.omprule} as a foundation, we introduce a submodular function that $(i)$ naturally tackles the aforementioned problem of OMP and $(ii)$ allows for near-optimal guarantees through Algorithm~\ref{GreedyAlgo1}.

\section{Sparse Representation and Submodularity}

In this section, we first discuss a closely related result in the intersection of sparse representation and submodularity. We then argue that if the \emph{average performance} of the reconstruction, over the input space, is considered, the resulting function can be shown to be submodular. We use this result to provide an efficient algorithm with probable guarantees in the next section.

\subsection{Sparse Representation Objective}
Let us consider a set of $m$ target signals, $\{\by^{(i)}\}_{i=1}^m$. Given that our goal is to best represent the target signals, we can consider the residual [c.f.~\eqref{eq.omprule}] for the last step, and put forth the following objective for constructing the solution set
\begin{align}\label{eq.objm}
    f_m(\mathcal{S}) := \frac{1}{m}\sum_{i}\big[\Vert \by^{(i)} \Vert_2^2 - \Vert \by^{(i)} - \by_{\mathcal{S}}^{(i)} \Vert_2^2\big], 
\end{align}
where $\by^{(i)}_\mathcal{S}$ denotes the projection of the $i$th target signal onto $\operatorname{span}\{\bPhi_{\mathcal{S}}\}$. 

The objective function~\eqref{eq.objm} has been considered before in~\cite{Krause3104395} for the dictionary selection problem and its \emph{approximate submodularity} has been shown. Here, we restate this result for completeness.

\begin{theorem}\cite{Krause3104395} If the dictionary $\bPhi$ has incoherence $\mu~:=~\underset{\forall (i,j),i\neq j}{\max}|\langle \bphi_i,\bphi_k\rangle|$, then $f_m$ satisfies
\begin{align}
    f_m(\mathcal{S}\cup\{a\}) - f_m(\mathcal{S}) \geq f_m(&\mathcal{S}\cup\{a,b\}) \\
    &- f_m(\mathcal{S}\cup\{a\}) -\epsilon
\end{align}
for all $a,b\in\mathcal{N}\setminus\mathcal{S}$; where $\epsilon \leq 4K\mu$.
\end{theorem}

Although the above result, in combination with the guarantees provided in~\cite{Krause3104395}, portraits an optimistic landscape for finding a \emph{near-optimal} solution to the dictionary learning problem (an instance of the sparse representation problem), the guarantees available for Algorithm~\ref{GreedyAlgo1} degrade with the number of desired atoms, $K$ and does not depend on the number, $m$, of considered target signals.

In spite of the above issue with the near-optimality guarantee, in practice, the greedy algorithm empirically performs well for large $m$. Here, we argue that this has to do with the properties of~\eqref{eq.objm} when $m\rightarrow\infty$.
\subsection{Asymptotics of Sparse Repreentation Objective Function}
Let the objective function for $m\rightarrow\infty$ be given by
\begin{equation}\label{eq.avgobj}
    f(\mathcal{S}) :=\lim_{m\rightarrow\infty}f_m(\mathcal{S}) = \mathbb{E}\{||\by||_2^2 - ||R_{\mathcal{S}}||_2^2\},
\end{equation}
where $\mathbb{E}\{\cdot\}$ stands for the expectation over the input space; and $R_\mathcal{S}:= \by-\by_{\mathcal{S}}$. 

After some mathematical manipulations (see Appendix~\ref{ap.margin}), the marginal gain of~\eqref{eq.avgobj}, i.e.,$\Delta(s|\mathcal{S}) := f(s\cup\mathcal{S}) - f(\mathcal{S})$, can be written as
\begin{align}\label{eq.delta2}
    \Delta(s|\mathcal{S})  = {\rm tr}\bigg\{\frac{1}{\Vert\bv_s\Vert_2^2}\bv_s^H\bR_\by\bv_s\bigg\}
\end{align}
where $\bv_s := (\bI - \bPi_\mathcal{S})\bphi_s$; and $\bR_{\by}:=\mathbb{E}\{\by\by^H\}$.

Note that when there is no structure in the input space, i.e., $\bR_{\by} = \bI$, the marginal gain $\Delta(s|\mathcal{S})$ is independent of both $\mathcal{S}$ and $s$; that is,
\begin{align}\label{eq.delta1}
    \Delta(s|\mathcal{S}) = 1\; \forall\; s \in\mathcal{N}\setminus\mathcal{S},\;\mathcal{S}\subset\mathcal{N}.
\end{align}

The assertion in~\eqref{eq.delta1} implies that in such a case,~\eqref{eq.avgobj} is a \emph{modular} set function, i.e., equality holds in~\eqref{equ:subcondorig}. 

In practice, a sparse representation is usually searched under the assumption that there exists some \emph{structure} in the input space, hence the above results seem irrelevant. However, building on the developed intuition, in the following, we show that function~\eqref{eq.avgobj} is indeed submodular \emph{independently} of the structure of the input signal space.

Before showing the submodularity of $f$, we establish its following property.

\begin{proposition} The set function $f$ is a normalized and nondecreasing (monotonic) set function.
\end{proposition}
\begin{proof}
We show that $f$ is normalized by noticing that if $\cal{S}=\emptyset$, then $\by_{\cal{S}}=\bzero$, and therefore $f(\emptyset) = \mathbb{E}\{||\by||_2^2 - ||\by||_2^2\} = 0$. To show the monotonic behavior of $f$, it is sufficient to show that $f(\mathcal{S})\leqslant f(\mathcal{S}\cup\{a\})$ where $a\notin\mathcal{S}$ is an index of a column of $\bPhi$. This fact can be seen by setting the $a$th coefficient of $\by_{\mathcal{S}\cup\{a\}}$ to $0$ (a special case of $\by_{\mathcal{S}\cup\{a\}}$); this selection leads to $f(\mathcal{S})= f(\mathcal{S}\cup\{a\})$. Thus, in general, $f(\mathcal{S})\leqslant f(\mathcal{S}\cup\{a\})$. 
\end{proof}


Now, to show the submodularity of $f(\cdot)$, we need to introduce the following result. 

\begin{lemma}
	\label{lemma1}
	Let $\by$ be a vector in a Hilbert space $\mathcal{H}$. If $\bu,\bv\in\mathcal{H}$ be two normalized vectors (i.e., unit norm) such that $\operatorname{span}(\bu,\bv)=\mathcal{H}$, then, the following inequality holds:
	\begin{equation}
	\label{equ:lemma1-1}
	\mathbb{E}\{||\by||_2^2\} \geqslant \mathbb{E}\{||\by - <\by,\bu>\bu||_2^2 + ||\by - <\by,\bv>\bv||_2^2\}.
	\end{equation}
\end{lemma}
\begin{proof}
See Appendix~\ref{prooflemma1}.
\end{proof}

The above lemma provides a way to guarantee that on average, the energy of the target signal is not over-represented by arbitrary vectors spanning the space wherein it lives. Using the above lemma, we are ready to state the following result.

\begin{theorem}
	The function $f(\cdot)$ defined in \eqref{eq.avgobj} is submodular. 	
	\label{theorem2}
\end{theorem}
\begin{proof}
See Appendix \ref{prooftheorem2}.
\end{proof}

The result in Theorem~\ref{theorem2} has the following implication.

\begin{corollary} Let $f$ be given as~\eqref{eq.avgobj}, and $\mathcal{I} = \mathcal{I}_K$, for some prescribed $K$. Then, the set $\mathcal{S}_{\rm G}$ returned by Algorithm~\ref{GreedyAlgo1} provides a $(1-{\rm e}^{-1})$-approximation for the cardinality-constrained subset selection problem.
\end{corollary}

 Despite that the above results guarantee near-optimality for the set constructed by the greedy algorithm, this, in general, requires knowledge of the statistics of the input space, i.e, $\bR_{\by}$. Unfortunately, in many cases, this information is not available (or is only known approximately). Thus, Algorithm~\ref{GreedyAlgo1} cannot be directly employed in practice. In the next section, we retake our previous toy example and discuss the single-point-estimate version of the greedy algorithm which deals with the (possible) lack of statistics of the input space. And we provide near-optimality guarantees for the finite-sample case of~\eqref{eq.avgobj}.

\renewcommand{\algorithmicrequire}{\textbf{Input:}}
\renewcommand{\algorithmicensure}{\textbf{Output:}}
\begin{algorithm}[t]\caption{\sc Greedy Algorithm}
	\label{GreedyAlgo1}
	\begin{algorithmic} [1]
		\REQUIRE $f(\cdot)$, $K$, $\mathcal{N}$.
		\ENSURE $\mathcal{S}_K$.
		
		{\bf Initialization:} $\mathcal{S}_0 = \emptyset$;	
				
		\FOR {$k = 1$ to $K$} 		
		\STATE $i^* = \underset{i\in \mathcal{N}\setminus\mathcal{S}_{k-1}}{\arg\max}~ f(\mathcal{S}\cup\{i\})$
		\STATE $\mathcal{S}_k \gets \mathcal{S}_{k-1}\cup \{i^*\}$;
		\ENDFOR		
	\end{algorithmic}
\end{algorithm}

\section{Submodular Matching Pursuit}
Recall that to maximize $f$ [c.f.~\eqref{eq.avgobj}], Algorithm~\ref{GreedyAlgo1} selects, at the $k$th step, the element which maximizes the marginal gain; that is,
\begin{align}\label{eq.smprule}
    i^* = \underset{i\not\in\mathcal{S}_{k-1}}{\arg\max}\; \Delta(i|\mathcal{S}_{k-1}) = \mathbb{E}\bigg\{\frac{\langle R_{\mathcal{S}_{k-1}},\bphi_i\rangle^2}{\Vert (\bI - \bPi_{\mathcal{S}_{k-1}})\bphi_i\Vert_2^2}\bigg\}.
\end{align}

Different to OMP (and MP), the greedy heuristic in~\eqref{eq.smprule} selects the column of the dictionary with the largest \emph{expected weighted} inner product with respect to the residual. The applied weights in~\eqref{eq.smprule} are given by the norm of the component of the column (atom) living in the orthogonal subspace to $\operatorname{span}\{\bPhi_{\mathcal{S}}\}$. From this point on, we refer to Algorithm~\ref{GreedyAlgo1} as Submodular Matching Pursuit (SMP) when the greedy rule~\eqref{eq.smprule} is employed.

\subsection{Revisiting Our Motivating Example}

To illustrate how the weights of~\eqref{eq.smprule} work, let us recall our toy example [c.f.~\eqref{equ:toy1}] and consider that the expectation in~\eqref{eq.smprule} is removed, and only a particular target signal is assumed. After the first step of the greedy algorithm, i.e., $\mathcal{S}_{1} = \{1\}$, we have the following \emph{reweighted} atoms
\begin{equation}
\mathcal{V}_1 := \bigg\{\tilde{\bv}_2:=\begin{bmatrix}
0 \\ 1 \\ 0
\end{bmatrix}, \tilde{\bv}_3:=\begin{bmatrix}
0 \\ 0 \\ 1
\end{bmatrix}\bigg\},
\label{equ:toy2}
\end{equation}
where the atoms are the \emph{decorrelated} versions of the previous columns, i.e., $\tilde{\bv}_i = (\bI - \bPi_{\{1\}})\bphi_i/\Vert (\bI - \bPi_{\{1\}})\bphi_i \Vert_2$. Notice that the last atom is kept unchanged as it has a zero inner product with the first column. Calculating the inner products with the target signal we have
\begin{equation}
\begin{aligned}
\langle \by,\tilde{\bv}_2\rangle &= 10, \\
\langle \by,\tilde{\bv}_3\rangle &= 1.
\end{aligned}
\end{equation}
Thus, the second column is selected in the second iteration. This version of the algorithm, referred as single-point-estimate SMP (see Algorithm~\ref{SOMP1}), is indeed the optimized orthogonal matching pursuit (OOMP) method~\cite{1001652}. Here, we refer to it as \emph{single-point-estimate} as the expectation is approximated with a single sample (unique target signal). Although OOMP is originally derived to alleviate the common issues of OMP, we arrive at it by means of submodularity; a completely different route. This further supports the evidence that OOMP can achieve good results in practice as it leverages intrinsically the submodularity of the underlying cost function.

\begin{algorithm}[t]\caption{\sc SMP :: Single-point-estimate}
	\label{SOMP1}
	\begin{algorithmic} [1]
		\REQUIRE $\bPhi$, $\by$, $K$.
		\ENSURE $\mathcal{S}$.
		
		{\bf Initialization:}
		$\mathcal{S} = \emptyset$;\;
		$R = \by$;\;
		$\bZ = []$;\;
		$\bV = []$\;		
		\FOR {$k = 1$ to $K$} 
		\FOR {$i\notin \mathcal{S}$}
		\IF {$k=1$}
		\STATE $\bPi_{\mathcal{S}} = {\bf 0}$\;
		\ENDIF
		\STATE $\tilde{\bphi}_i =  (\bI - \bPi_{\mathcal{S}})\bphi_i$
		\STATE $\tilde{\bphi}_i = {\tilde{\bphi}_i}/{||\tilde{\bphi}_i||_2}$
		\ENDFOR
		\STATE $i^* = {\arg\max}_{i\notin \mathcal{S}}\;|\tilde{\bphi}_i^H R|$\;
		\STATE $\mathcal{S} = \mathcal{S}\cup \{i^*\}$\;
		\STATE $\bV = [\bV\, \tilde{\bphi}_{i^*}]$
		\STATE Update $\bZ = (\bV^H\bV)^{-1}$ using \eqref{equ:inverse2}
		\STATE $\bPi_{\mathcal{S}} = \bV \bZ \bV^H$\
		\STATE $R = (\bI - \bPi_{\mathcal{S}})\by$
		\ENDFOR		
	\end{algorithmic}
\end{algorithm}

\subsection{A Finite-Sample Result for Submodular Matching Pursuit }

Though we have shown that asymptotically, $m\rightarrow\infty$, the set function $f_m$ is a submodular set function, most of the cases of practical interest occur when $m$ is finite; that is, in the \emph{finite-sample regime}. Thus, in the following, we provide a near-optimality guarantee for the case where the finite-sample version [c.f.~\eqref{eq.objm}] (and therefore any variant of SMP) is used to find a solution for~\eqref{eq.avgobj} using Algorithm~\ref{GreedyAlgo1}.

\begin{theorem} Let $f_m$ and $f$ be defined as in~\eqref{eq.objm} and~\eqref{eq.avgobj}, respectively. Further, assume that the target signals $\{\by^{(i)}\}_{i=1}^{m}$ are i.i.d. Given a prescribed $K$ and tolerance $\epsilon_m$, the set $\mathcal{S}_{\rm G}$ returned by Algorithm~\ref{GreedyAlgo1}, using $f_m$ as input, satisfies
\begin{equation}
    {\rm P}(f_m(\Sc_{\rm G}) \geq (1 - {\rm e}^{-1})f(\So) - (2K+1)\epsilon_m) = {p_K}
\end{equation}
    where 
\begin{align}
    \ln p_K = (2K+1)K\ln(1 - \sigma^2/m\epsilon_m^2),
\end{align}    
    where $\sigma^2 ~:=~\max_{\mathcal{S}\subseteq V}\mathbb{E}\{|f_m(\mathcal{S}) - f(\mathcal{S})|^2\}$.
    \label{theoremLast}
\end{theorem}

\begin{proof}
 See Appendix~\ref{ap.prooflast}
\end{proof}

This result is similar in flavor to that of~\cite{Krause3104395} in the sense that it shows the \emph{near-optimality} of employing the cost function~\eqref{eq.objm} to solve the signal representation problem, but differs in the following aspects: $(i)$ it is probabilistic; $(ii)$ it has a dependency on the number of target signals considered; and $(iii)$ it is applicable to Algorithm~\ref{SOMP1} even though the true objective to maximize is $f$. Due to $(iii)$, we can then guarantee (probabilistically) the near-optimality of the set returned by Algorithm~\ref{SOMP1} (and hence OOMP) for the signal representation problem as it is a special case of the result of Theorem~\ref{theoremLast} with $m=1$.

\begin{figure*}[h!]
	\centering
	\psfrag{Original}{\scriptsize{Original}}
	\psfrag{Exhaustive Search}{\scriptsize{Exhaustive Search}}
	\psfrag{SMP}{\scriptsize{SMP}}
	\psfrag{OMP}{\scriptsize{OMP}}
	\psfrag{MP}{\scriptsize{MP}}
	\psfrag{Objective Function}{{Objective Function}}
	\psfrag{Estimation Error}{{Estimation Error}}
	\psfrag{K}{{$K$}}
	\subfigure[] {\includegraphics[width=.48\textwidth]{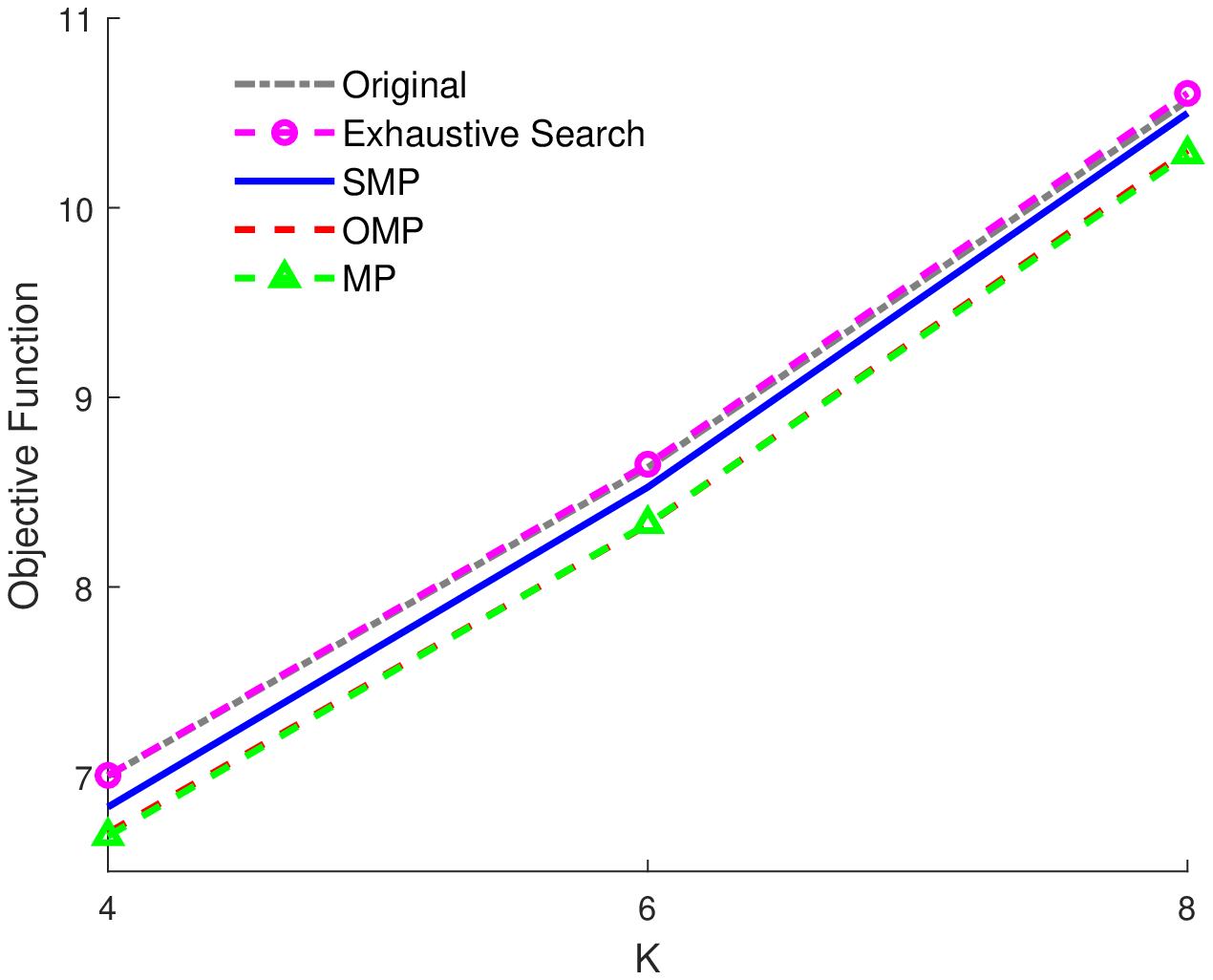}
		\label{card_small_obj}} \quad
	\subfigure[] {\includegraphics[width=.48\textwidth]{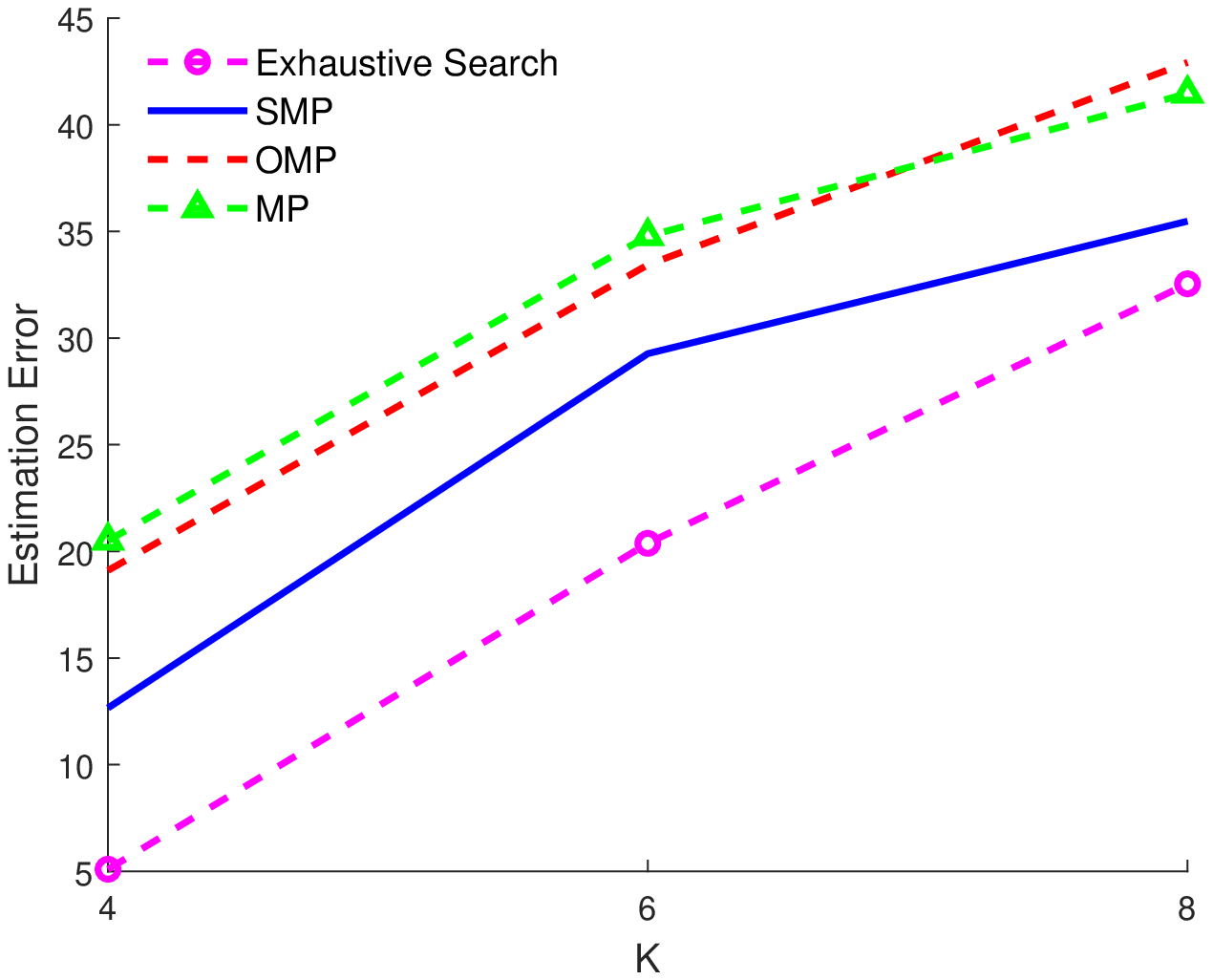}
		\label{card_small_est}} \\
	\caption{Performance comparison of different algorithms for a small scenario with a cardinality constraint (a) Objective function versus $K$, (b) Estimation error versus $K$.}
	\label{card_small}
\end{figure*}

\section{Constrained Submodular Matching Pursuit}

Although in the literature there have been efforts to extend classical matching pursuit algorithms to settings with constraints, see, e.g.,~\cite{adler2011constrained,huang2012adaptive, yaghoobi2015fast, baraldi2019basis}, most of this work either focus on linear constraints, i.e., restriction of the solution to a linear subspace, or in the modification altogether of the cost to optimize to promote structural characteristics in the solution. These considerations lead to algorithms, that though based on MP methods, deviate significantly from their original formulation. Thus, in many cases, they do not enjoy the simplicity of MP approaches. 

Deviating from the type of constraints addressed in the above-mentioned works, we here shift our attention to constraints that are pervasive in selection problems: \emph{combinatorial constraints}. And we show that for these kinds of constraints, under minor modifications to Algorithm~\ref{SOMP1}, near-optimal guarantees can be provided. In this section, we first formally introduce these kinds of constraints, namely knapsack and matroid constraints, and provide examples of applications, within signal processing, wherein they naturally appear. Then, we discuss the modifications to Algorithm~\ref{SOMP1} that are required to provide the near-optimal guarantees by means of a greedy-heuristic.

\subsection{Knapsack Constraints}
Instead of enforcing a cardinality constraint for the subset selection problem in Definition 3, in many applications, the elements in $\mathcal{N}$ might have non-uniform costs, i.e., $p_s, s\in\mathcal{N}$. Thus, given a budget $B$, we are interested in finding a set $\mathcal{S}\subseteq \mathcal{N}$ maximizing $f(\mathcal{S})$, subject to the budget constraint $p_{\mathcal{S}}=\sum_{s\in\mathcal{S}}p_s\leqslant B$. 
{This situation arises, for example, in the sensing coverage problem (Figure \ref{warmup_example}) where each of the sensors has a given operative cost.  Different sources of energy, variety of accommodation costs, or level of energy efficiency can be the reason behind this difference in operative cost. Subsequently, in such cases, the constraint on the number of sensors, the cardinality constraint, converts to a budget constraint in which the total cost of selected sensors should not exceed the given budget.}

In order to maximize the knapsack constrained representation problem, we build a solution using two variants of the proposed SMP algorithm. The first variant substitutes the selection rule (line $9$) in Algorithm~\ref{SOMP1} by 
\begin{align}\label{eq.xmprule}
    i^* = {\arg\max}&\;\;|\tilde{\bphi}_i^H R|,\\
    {\rm s.t.} &\;\; i\in\{j : j\notin \mathcal{S},  \,p_{\mathcal{S}\cup\{j\}}\leq B\}\nonumber,
\end{align}
and in the second variant, the selection rule is substituted by the following rule
\begin{align}\label{eq.xmprule2}
    i^* = {\arg\max}&\;\;{|\tilde{\bphi}_i^H R|}/p_i\\
    {\rm s.t.} &\;\; i\in\{j : j\notin \mathcal{S},  \,p_{\mathcal{S}\cup\{j\}}\leq B\}\nonumber.
\end{align}
Both algorithms stop when there is no element satisfying the constraints. It is shown that if $f$ is a normalized, monotonic, submodular set function, executing both variants of the proposed algorithm and selecting the best of the two results, achieves the optimal solution with the factor $\frac{1}{2}(1-\rm{e}^{-1})$ \cite{leskovec2007cost}.
Further, with a partial enumeration of all feasible sets of cardinality one or two, and substituting the selection rule \eqref{eq.xmprule2} in Algorithm~\ref{SOMP1}, a $(1-1/\rm{e})$-approximation guarantee is achieved \cite{sviridenko2004note}.

\subsection{Matroid Constraints}
A matroid is defined as a pair $(\mathcal{N}, \mathcal{I})$ in which $\mathcal{N}$ is a finite set and $\mathcal{I}\subseteq 2^{\mathcal{N}}$ comprises any subset of $\mathcal{N}$ which satisfies the following properties: 
\begin{itemize}
	\item $\mathcal{A}\subseteq \mathcal{B}\subseteq \mathcal{N}$ and $\mathcal{B}\in \mathcal{I}$ implies $\mathcal{A}\in \mathcal{I}$.
	\item $\mathcal{A},\mathcal{B}\in \mathcal{I}$ and $|\mathcal{B}|>|\mathcal{A}|$ implies that $\exists a \in \mathcal{B}\,\backslash\,\mathcal{A}$ such that $\mathcal{A}\cup\{a\}\in\mathcal{I}$.
\end{itemize}
{Recalling the sensing coverage problem, in some applications, the existence of some of the sensors can be mutually exclusive. For example, in case the location of candidate sensors are physically overlapping, only one would fit in the considered position.}
{Also, one can consider the case that different subgroups of sensors are fed from particular sources of energy with limited capacities. This could lead, for instance, to constraints with respect to the number of selected sensors from the different subgroups.
To provide a more concrete example, let us consider $\{\mathcal{N}_i, i=1,...,M\}$ to be a partition of the ground set $\mathcal{N}$, i.e., $\mathcal{N}= \bigcup_{i=1}^M \mathcal{N}_i:\mathcal{N}_i\cap\mathcal{N}_j=\emptyset\;\forall\;i,j$. Defining the matroid $(\mathcal{N},\mathcal{I})$ with $\mathcal{I} = \{\mathcal{S}|\mathcal{S}\subseteq \mathcal{N}, |\mathcal{S}\cap \mathcal{N}_i|\leqslant K_i, i=1,...,M\}$,  we obtain a so-called \emph{partition matroid}. This kind of matroids has applications, for instance, in the problem of joint transmit/receive antenna selection, e.g.,~\cite{8537943}. In this setting, the partition matroid, $(\mathcal{N},\mathcal{I})$, is composed by the sets of transmitters and receivers, and $K_1$ and $K_2$ denote the maximum allowed number of transmitters and receivers, respectively, that can be selected.}

Recalling the subset selection problem in Definition 3, to maximize the objective function subject to a given matroid $(\mathcal{N}, \mathcal{I})$, we employ the proposed SMP algorithm by substituting the selection rule (line $9$) in Algorithm~\ref{SOMP1} by
\begin{align}\label{eq.xmprule3}
    i^* = {\arg\max}&\;\;|\tilde{\bphi}_i^H R|,\\
    {\rm s.t.} &\;\; i\in\{j : j \notin \mathcal{S},  \,\mathcal{S}\cup\{j\}\in\mathcal{I}\}\nonumber.
\end{align}
and the algorithm stops when there is no such an element. It is shown that if $f$ is a normalized, monotonic, submodular set function, the proposed algorithm achieves the optimal solution with the factor $\frac{1}{2}$ \cite{fisher1978analysis}. Alternatively, a $(1-1/\text{e})$-approximation guarantee can be achieved if the continuous greedy algorithm in \cite{calinescu2011maximizing} is used instead.

\begin{figure*}[h]
	\centering
	\psfrag{Original}{\scriptsize{Original}}
	\psfrag{Exhaustive Search}{\scriptsize{Exhaustive Search}}
	\psfrag{SMP}{\scriptsize{SMP}}
	\psfrag{OMP}{\scriptsize{OMP}}
	\psfrag{MP}{\scriptsize{MP}}
	\psfrag{Objective Function}{{Objective Function}}
	\psfrag{Estimation Error}{{Estimation Error}}
	\psfrag{K}{{$K$}}
	\subfigure[] {\includegraphics[width=.48\textwidth]{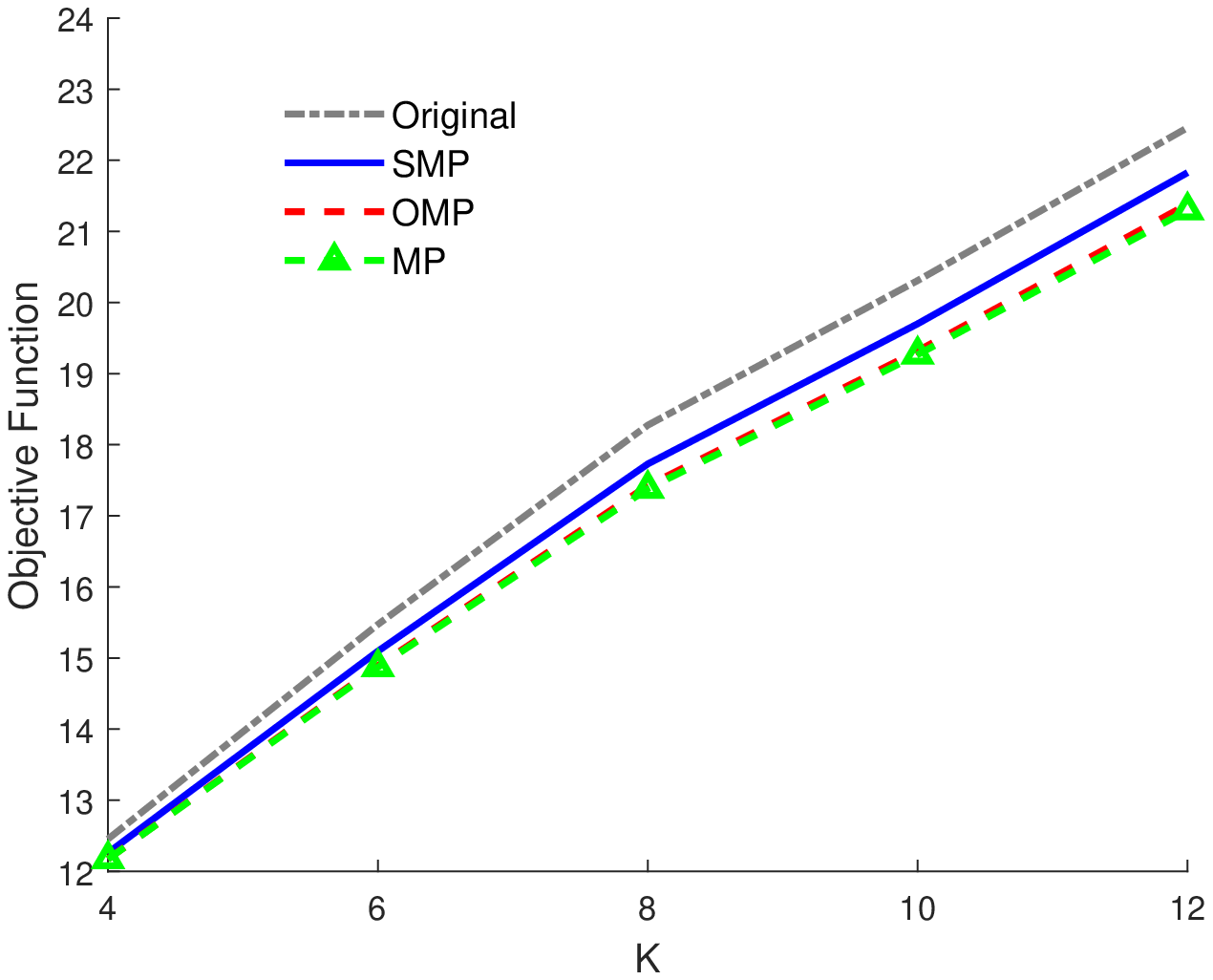}
		\label{card_large_obj}} \quad
	\subfigure[] {\includegraphics[width=.48\textwidth]{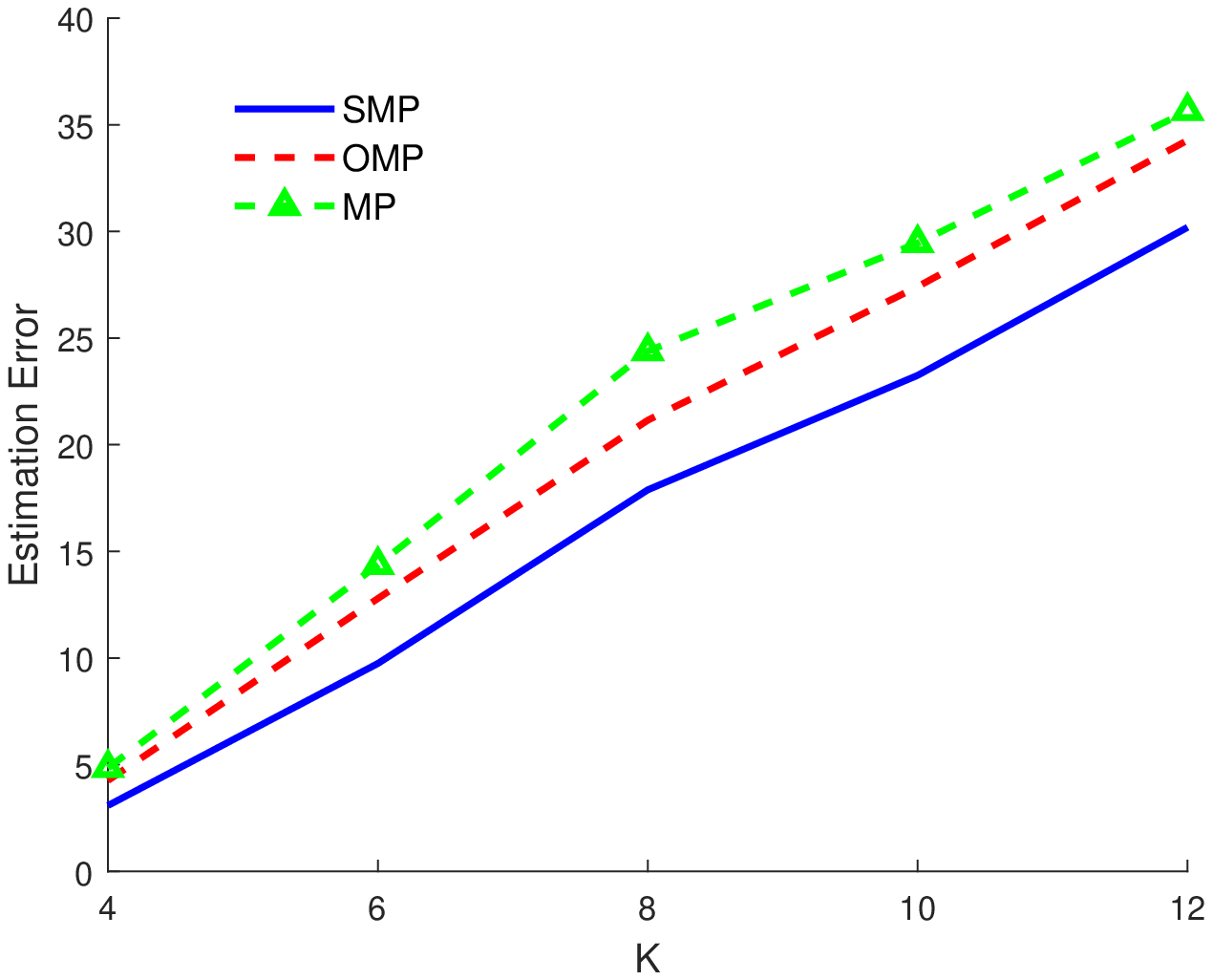}
		\label{card_large_est}} \\
	\caption{Performance comparison of different algorithms for a large scenario with a cardinality constraint (a) Objective function versus $K$, (b) Estimation error versus $K$.}
	\label{card_large}
\end{figure*}
\begin{figure*}[h]
	\centering
	\psfrag{Original}{\scriptsize{Original}}
	\psfrag{Exhaustive Search}{\scriptsize{Exhaustive Search}}
	\psfrag{SMP}{\scriptsize{SMP}}
	\psfrag{OMP}{\scriptsize{OMP}}
	\psfrag{MP}{\scriptsize{MP}}
	\psfrag{Objective Function}{{Objective Function}}
	\psfrag{Estimation Error}{{Estimation Error}}
	\psfrag{K}{{$K$}}
	\subfigure[] {\includegraphics[width=.48\textwidth]{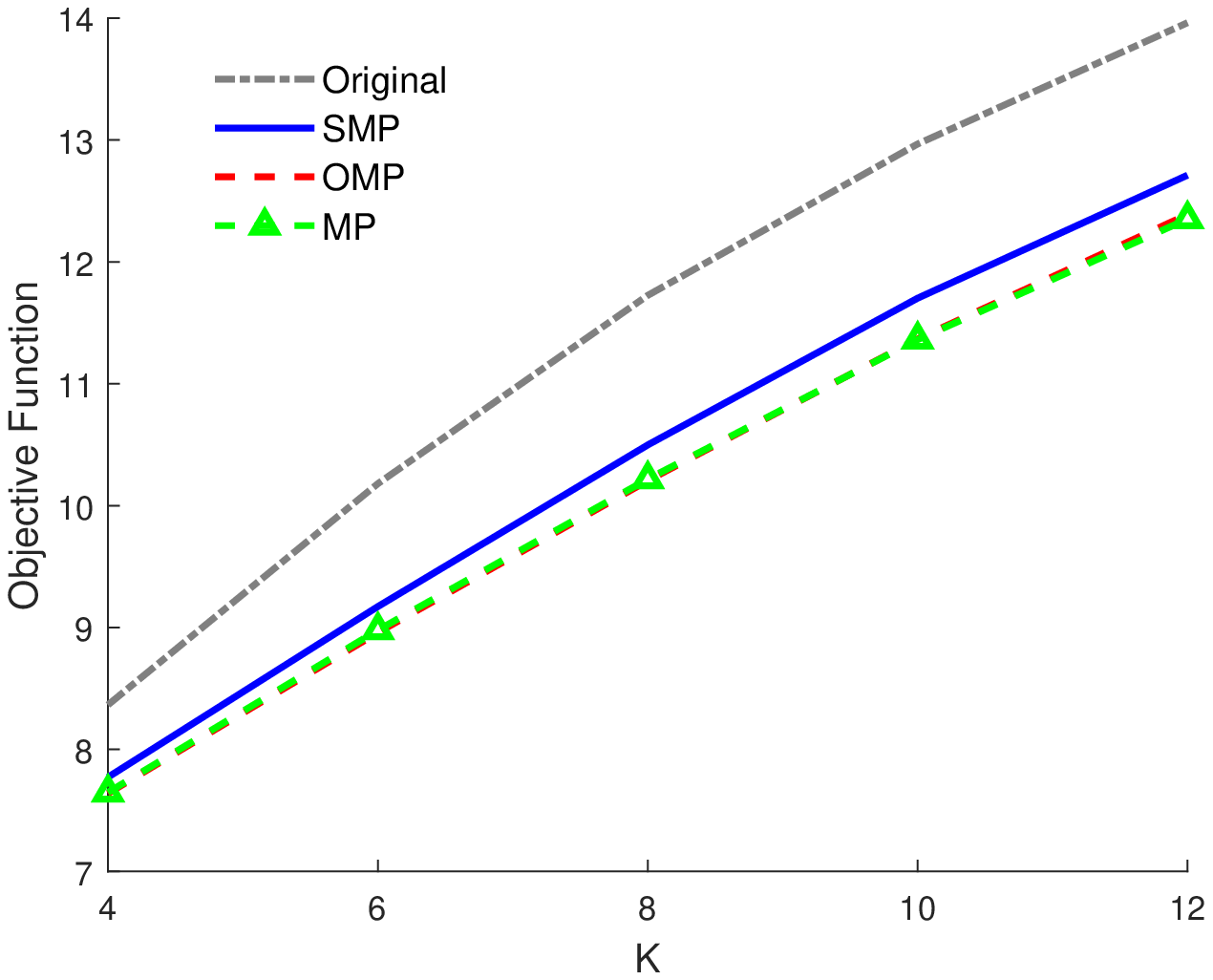}
		\label{mat_large_obj}} \quad
	\subfigure[] {\includegraphics[width=.48\textwidth]{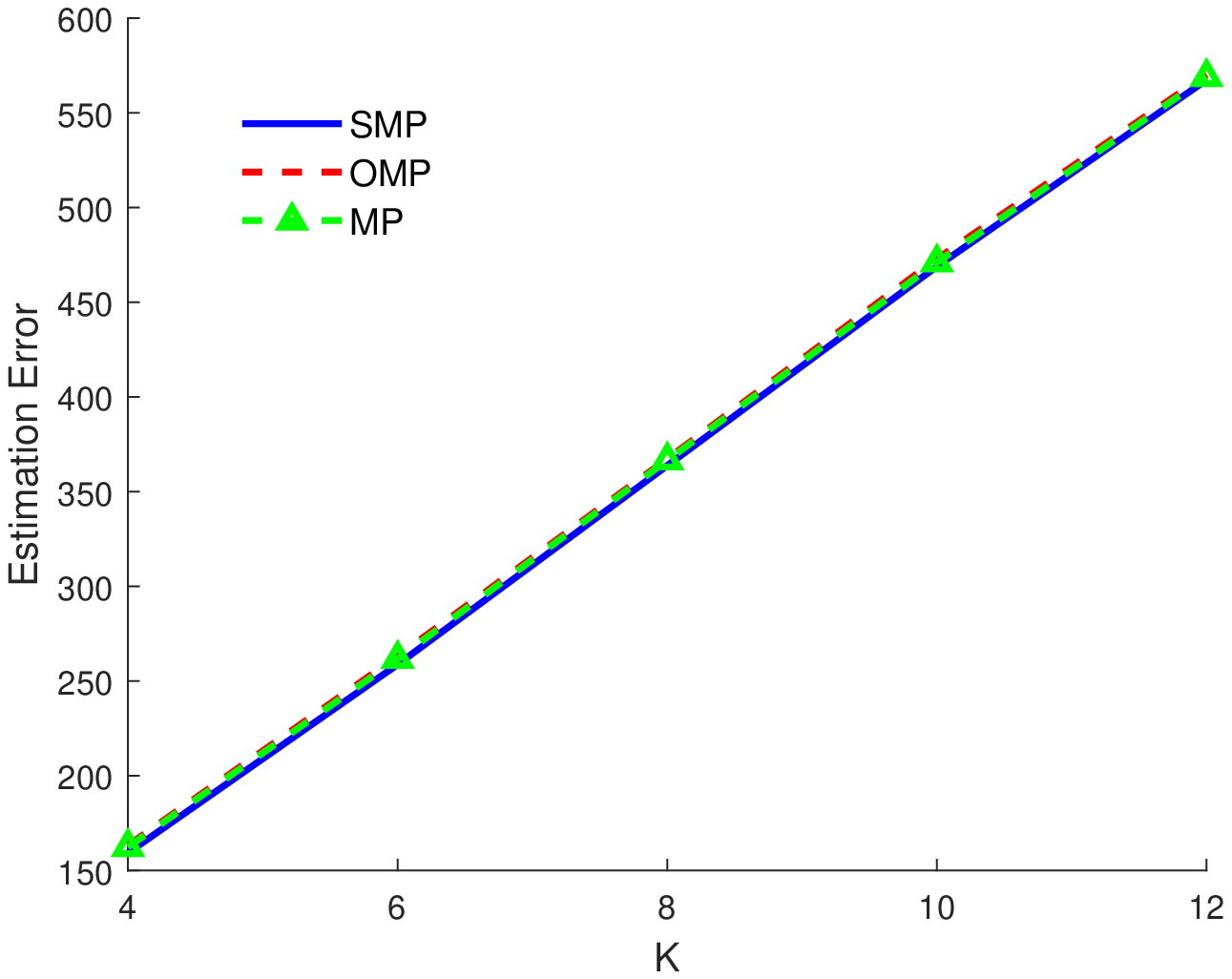}
		\label{mat_large_est}} \\
	\caption{Performance comparison of different algorithms for a large scenario with a matroid constraint (a) Objective function versus $K$, (b) Estimation error versus $K$.}
	\label{mat_large}
\end{figure*}
\section{Experiments}
In this section, we make a comparison between the SMP algorithm and the MP and OMP algorithms. First of all, the MP and OMP algorithms have computational complexities equal to $O(KMN)$ and $O(KMN+K^3)$, respectively. Comparing the algorithms in terms of computational complexity, the proposed SMP algorithm approximately requires $K$ times more computations. In the sequel, we compare the algorithms from their performance perspective.

For the following set of simulations, we consider the direction of arrival (DOA) estimation problem~\cite{van2004optimum}. We assume an array of $M$ sensors placed along a line with $\lambda/2$ spacing ($\lambda$ is the wavelength). Moreover, we grid the total angle span of interest, for these simulations $[-80,80]$ degree, into a set of $N$ discrete angle cells. Using the formulation in \eqref{equ:recovery}, we consider an additive white Gaussian noise (AWGN) with a signal to noise ratio (SNR) equal to $20$dB. Also, we consider there exist $K$ sources in the assumed angle span. To evaluate the performance of the algorithms for both signal representation and recovery problems, we consider two metrics. One is the objective function introduced in \eqref{eq.avgobj} (without the expectation) which determines the quality of representation. The other metric is the estimation error which is defined as $||\btheta(\mathcal{S})-\btheta_{Original}||_2^2$ where $\btheta(\mathcal{S})$ and $\btheta_{Original}$ are the estimate and real DOAs, respectively.

For the first case, we compare SMP, OMP, and MP algorithms for a scenario with $M=10$ and $N=15$ for the problem with cardinality constraint. Since it is a small scenario, the optimal solution (the result of an exhaustive search) is also compared. It should be noted that since we are working with a noise-contaminated signal, the exhaustive search may result in a higher objective value in comparison with the original set of columns. Figure \ref{card_small_obj} depicts the objective function versus the cardinality constraint $K$. As observed in the figure, SMP outperforms OMP and MP, and approaches the optimal solution. Using the solution of these algorithms for signal representation, in Figure \ref{card_small_est}, the estimation error versus $K$ is plotted (clearly the estimation error for the original set of columns is zero and is not plotted). It is worthy to point out that a higher value of the objective function does not necessarily result in a lower estimation error. As shown in Figure \ref{card_small_est}, the SMP has a lower estimation error in comparison with OMP and MP.

For the second case, in Figure \ref{card_large}, we compare different algorithms for a scenario with $M=30$ and $N=100$ where we have the cardinality constraint. Again, the superiority of SMP over OMP and MP is clear in figures \ref{card_large_obj} and \ref{card_large_est}.

For the final case, we consider the representation problem subject to a matroid constraint. In this scenario, we have $M=30$ and $N=100$. Moreover, the matroid constraint is constructed as follows: we split the total 100 angle cells into 50 groups of non-overlapping angle cells each consists of 2 consecutive angle cells. We put the limitation that at most one angle cell from each group can be selected. This defines a partition matroid. Figure \ref{mat_large} depicts the results of different algorithms. In Figure \ref{mat_large_obj}, although the gap between the original solution and the result of SMP is increased, SMP is still better than OMP and MP algorithms. In addition, the estimation error plotted in Figure \ref{mat_large_est} presents similar results for all three algorithms.

Even though the results look quite good for cardinality constraints, the gains seem to be less impressive for matroid constraints, especially in Figure \ref{mat_large_est}. An intuitive explanation is that in the matroid example, we have two possibilities in each partition (consecutive angle cells). Even if the selected cell within the two cells of a partition is wrong, its contribution to the estimation error, i.e., $||\btheta(\mathcal{S})-\btheta_{Original}||_2^2$, is small, while the superiority of the SMP is still evident in Figure \ref{mat_large_obj} due to applying a different objective function.

\section{Conclusion}
Investigating the problem of signal representation from a new perspective, we introduced a new formulation and proved that the objective function is submodular. We discussed how the problems of signal representation and signal recovery are related, and thus, the same model and algorithm works for both. Besides the conventional cardinality constraint, for the first time, we posed this problem with a matroid constraint which enables using the proposed SMP algorithms for a wider variety of applications. Leveraging submodularity, we proposed the SMP algorithm that not only (for the first time) provides a valid near-optimal performance bound for any signal representation problem, but also resolves the existing deficiency in the OMP algorithm. In addition, we show the connection of the single-point-estimate version of SMP with OOMP. Finally, as shown through the simulations, the SMP algorithm outperforms the MP and OMP algorithms.

\appendices
\section{Marginal gain expression}\label{ap.margin}

Let us first expand the set function and rewrite it as
\begin{align}
    f(\Sc) &= \mathbb{E}\{\Vert \by \Vert_2^2 - \Vert \by - \by_{\Sc} \Vert_2^2\}  \\
     &= 2\mathbb{E}\{\langle \by,\by_\mathcal{S}\rangle\}- \mathbb{E}\{ \Vert \by_\mathcal{S} \Vert_2^2 \} \\
     &= 2\mathbb{E}\{\by^H\bPi_\mathcal{S}\by\} - \mathbb{E}\{(\by^H\bPi_\mathcal{S})(\bPi_\mathcal{S}\by)\}\\
     & = \mathbb{E}\{\by^H\bPi_{\mathcal{S}}\by\},
\end{align}
where the fact $\bPi_{\Sc}\bPi_{\Sc} = \bPi_{\Sc}$ has been used. Now, using the above expression, the marginal gain $\Delta(s|\mathcal{S})$ can be expressed as
\begin{align}
    \Delta(s|\mathcal{S}) &= f(\mathcal{S}\cup\{s\}) - f(\mathcal{S})\\
    &= \mathbb{E}\{\by^H(\bPi_\mathcal{S} +\bPi_{\tilde{\bv}_{s}})\by\} - \mathbb{E}\{\by^H\bPi_\mathcal{S}\by\}\\
    &= \mathbb{E}\{\by^H\bPi_{\tilde{\bv}_{s}}\by\}\label{eq.deltas}.
\end{align}
Recalling that $\tilde{\bv}_s:= \bv_s/\Vert\bv_s\Vert_2$, with $\bv_{s}:=(\bI-\bPi_{\mathcal{S}})\bphi_s$, the projection matrix $\bPi_{\tilde{\bv}_s}$ can be written as
\begin{align}
    \bPi_{\tilde{\bv}_s} &= \tilde{\bv}_{s}\tilde{\bv}_{s}^H
    = \frac{1}{\Vert\bv_s\Vert_2^2}\bv_s\bv_s^H.
\end{align}
Hence, applying trace properties to the terms in~\eqref{eq.deltas}, i.e.,
\begin{align}
    \Delta(s|\mathcal{S}) &= \mathbb{E}\{\by^H\bPi_{\tilde{\bv}_{s}}\by\}\\
    &= \mathbb{E}\{{\rm tr}\big(\by^H\bPi_{\tilde{\bv}_{s}}\by\big)\}\\
    &= {\rm tr}\big(\bPi_{\tilde{\bv}_{s}}\mathbb{E}\{\by\by^H\}\big),
\end{align}
and setting $\bR_{\by}:=\mathbb{E}\{\by\by^H\}$, the expression~\eqref{eq.delta2} for the marginal gain is obtained.

\section{Proof of Lemma \ref{lemma1}}
\label{prooflemma1}
\begin{proof}
	Expanding right side of \eqref{equ:lemma1-1} leads to
	\begin{equation}
	\begin{aligned}		
	&\mathbb{E}\{||\by||_2^2\} \geqslant \mathbb{E}\{||\by||_2^2 + <\by,\bu>^2 - 2<\by,\bu>^2 \\&+ ||\by||_2^2 + <\by,\bv>^2 - 2<\by,\bv>^2\} \\
	\Rightarrow \quad & \mathbb{E}\{<\by,\bu>^2 + <\by,\bv>^2\} \geqslant \mathbb{E}\{||\by||_2^2\}\\
	\Rightarrow \quad & \mathbb{E}\{\frac{||\by||_2^2}{2} + \frac{||\by||_2^2}{2}\} \geqslant \mathbb{E}\{||\by||_2^2\},
	\end{aligned}
	\label{equ.proof.lemma.1}
	\end{equation}
	where the last inequality is obtained by the following integration:
	\begin{equation}
	\begin{aligned}
	\mathbb{E}&\{<\by,\bu>^2\} = \frac{\mathbb{E}\{||\by||_2^2\}}{2\pi}\int_{0}^{2\pi}\cos^2\theta d\theta \\ &= \frac{\mathbb{E}\{||\by||_2^2\}}{2\pi}\int_{0}^{2\pi}\frac{1+\cos2\theta}{2} d\theta = \frac{\mathbb{E}\{||\by||_2^2\}}{2},
	\end{aligned}
	\label{equ.uniform.dist.proof}
	\end{equation}
	where a uniform distribution is considered for the angle between $\by$ and $\bu$. 
	This assumption is due to the fact that $\by$ in Lemma \ref{lemma1} is not the original signal but a projection of the original signal into the span of two vectors as derived in Theorem \ref{theorem2} and particularly in \eqref{equ:theorem-2}. As there is no prior information on these two vectors, it is reasonable to consider a uniform distribution for the angle between the projected part and $\bu$.
	
	It is worth mentioning that the uniform distribution is not the only distribution that brings the result of \eqref{equ.uniform.dist.proof}.
	Based on \eqref{equ.uniform.dist.proof}, it is straightforward to see that \eqref{equ.proof.lemma.1} is met for any probability density function (pdf) of the angle between $\by$ and $\bu$, denoted by $f(\theta)$, as long as the inequality $\int_{0}^{2\pi}f(\theta)\cos2\theta d\theta \geqslant 0$ is satisfied. Apart from the pdfs that hold the constraint in an equality form (e.g., the uniform distribution as shown in \eqref{equ.uniform.dist.proof}), for any pdf $f_1(\theta)$ that strictly satisfies the inequality, there is a corresponding pdf $f_2(\theta)=f_1((\theta+\pi/2)\mod 2\pi)$ that violates the constraint, and vice versa, i.e.,
	\begin{equation}
	\begin{aligned}
	    \int_{0}^{2\pi}f_2(\theta)\cos2\theta d\theta &= \int_{0}^{2\pi}f_1(\theta)\cos(2\theta-\pi) d\theta \\ &= -\int_{0}^{2\pi}f_1(\theta)\cos2\theta d\theta.
	    \end{aligned}
	\end{equation}
	Consequently, more than half of the possible pdfs (i.e., including the ones that satisfy in the equality form plus half of the rest) lead to the result of Lemma \ref{lemma1}.
\end{proof}

\section{Proof of Theorem \ref{theorem2}}
\label{prooftheorem2}
\begin{proof}
	Substituting \eqref{eq.avgobj} in \eqref{equ:subcondorig} leads to
	\begin{equation}
	\begin{aligned}
	&\mathbb{E}\{||\by||_2^2 - ||\by-\by_{\mathcal{S}\cup\{a\}}||_2^2 - ||\by||_2^2 + ||\by-\by_{\mathcal{S}}||_2^2\} \geqslant\\
	&\mathbb{E}\{||\by||_2^2 - ||\by-\by_{\mathcal{S}\cup\{a,b\}}||_2^2 - ||\by||_2^2 + ||\by-\by_{\mathcal{S}\cup\{b\}}||_2^2\}.
	\label{equ:sub1}
	\end{aligned}
	\end{equation}
	Simplifying equal terms from both sides, \eqref{equ:sub1} reduces to
	\begin{equation}
	\begin{aligned}
	&\mathbb{E}\{||\by-\by_{\mathcal{S}}||_2^2 - ||\by-\by_{\mathcal{S}\cup\{a\}}||_2^2\} \\&\geqslant
	\mathbb{E}\{||\by-\by_{\mathcal{S}\cup\{b\}}||_2^2 - ||\by-\by_{\mathcal{S}\cup\{a,b\}}||_2^2\}.
	\label{equ:sub2}
	\end{aligned}
	\end{equation}	
	Here, we introduce $\bphi_{\{\tilde{a}\}}$ and $\bphi_{\{\tilde{b}\}}$ in order to present the part of $\bphi_{\{a\}}$ and $\bphi_{\{b\}}$, respectively, that are not in $\operatorname{span}\{\bPhi_{\mathcal{S}}\}$, i.e., $\bphi_{\{\tilde{a}\}},\bphi_{\{\tilde{b}\}}\perp \operatorname{span}\{\bPhi_{\mathcal{S}}\}$.
	Moreover, we restate $\by$ as
    \begin{equation}
    \by=\tilde{\by}+\by_{\mathcal{S}}+\by_{\{\tilde{a},\tilde{b}\}}, 
    \label{equ:zbreak}
    \end{equation}
    where $\tilde{\by}$ is defined as the residual of projecting $\by$ over $\operatorname{span}\{\bPhi_{\mathcal{S}\cup\{\tilde{a},\tilde{b}\}}\}$, i.e., $\tilde{\by} = \by -\by_{\mathcal{S}\cup\{\tilde{a},\tilde{b}\}}$.
	Expanding $\by$ using \eqref{equ:zbreak} and following some simplifications we obtain
	\begin{equation}
	\begin{aligned}
	&\mathbb{E}\{||\tilde{\by}+\by_{\{\tilde{a},\tilde{b}\}}||_2^2 - ||\tilde{\by}+\by_{\{\tilde{a},\tilde{b}\}} - \by_{\{\tilde{a}\}}||_2^2\} \\&\geqslant \mathbb{E}\{||\tilde{\by}+\by_{\{\tilde{a},\tilde{b}\}}-\by_{\{\tilde{b}\}}||_2^2 - ||\tilde{\by}||_2^2\}.		
	\end{aligned}
	\label{equ:theorem-1}
	\end{equation}
	Since $\tilde{\by}$ is orthogonal to the other terms in \eqref{equ:theorem-1}, the following inequality is obtained:
	\begin{equation}
	\begin{aligned}
	&\mathbb{E}\{||\by_{\{\tilde{a},\tilde{b}\}}||_2^2\} \\ &\geqslant \mathbb{E}\{||\by_{\{\tilde{a},\tilde{b}\}} - \by_{\{\tilde{a}\}}||_2^2 + ||\by_{\{\tilde{a},\tilde{b}\}}-\by_{\{\tilde{b}\}}||_2^2\},	
	\end{aligned}
	\label{equ:theorem-2}
	\end{equation}
	which is correct based on Lemma \ref{lemma1}.
\end{proof}

\section{Proof of Theorem \ref{theoremLast}}\label{ap.prooflast}
\begin{proof}
To show the result, consider the following chain of inequalities, i.e.,
\begin{align}
    f(\So) & \overset{(a)}{\leq} f(\So \cup \Si) \\
           & \overset{(b)}{=} f(\Si) + \sum_{j=1}^{k} \Delta(v_j^*|\Si\cup v_1,\ldots,v_{j-1}) \\
           &\overset{(c)}{\leq} f(\Si) + \sum_{v\in\Sc^*}\Delta(v|\Si) \\
           & \overset{(d)}{\leq} f_m(\Si) + \epsilon_m +  \sum_{v\in\So}[\Delta_{i+1} + 2\epsilon_m] \\
           & = f_m(\Si) + K\Delta_{i+1} + \alpha\epsilon_m,
\end{align}
where $(a)$ is due to monotonicity; $(b)$ is established by a telescopic sum; $(c)$ is result of submodularity; and $(d)$ holds with probability $(1 - \sigma^2/m\epsilon_m^2)^{2K+1}$ by a Chernoff's bound-type inequality. $\Delta_{i+1}:= f_m(\Sc_{i+1}) - f_m(\Si)$ and $\alpha = 2K + 1$.

Now, we can define the following quantity.
\begin{align}
    \delta_i & := f(\So) - f_m(\Si) \\
    & \overset{(a)}{\leq} K\Delta_{i+1} + \alpha\epsilon_m \\
    &  = K(\delta_i - \delta_{i+1}) + \alpha\epsilon_m,
\end{align}
where $(a)$ is established using the previous inequalities. Rearranging terms from the above expression, we get
\begin{align}
     \delta_{i+1} \leq \frac{(K - 1)}{K}\delta_{i} + \epsilon_m',
\end{align}
where $\epsilon_m' := (2 + K^{-1})\epsilon_m$.

Finally, applying the above inequality repetitively ($k$ times), we obtain
\begin{align}
    \delta_k & \leq \bigg(\frac{K-1}{K}\bigg)^{k}\delta_0 + \epsilon_m'\sum\limits_{l=0}^{k-1} \bigg(\frac{K-1}{K}\bigg)^{l} \\
    & \overset{(a)}{\leq} {\rm e}^{-k/K}f(\So) + k\epsilon_m',
\end{align}
which leads to the desired result
\begin{align}
    f_m(\Sc_{\rm G}) \geq (1 - {\rm e}^{-1})f(\So) - K\epsilon_m'.
\end{align}

\end{proof}

\section{Computational Aspects: Matrix Update}

Given a matrix $\bX$ with inverse $\bY = (\bX^H\bX)^{-1}$, the procedure for updating $\bY$, when a column is added to the matrix $\bX$, is given by the following procedure.

Consider $\bX_{\rm new} = \begin{bmatrix} 
\bX & \bv
\end{bmatrix}$, thus
\begin{equation}
\begin{aligned}
\bY_{\rm new}^{-1} &= \bX_{\rm new}^H\bX_{\rm new} = \begin{bmatrix} 
\bX^H \\ \bv^H
\end{bmatrix}\begin{bmatrix} 
\bX & \bv
\end{bmatrix} \\&= \begin{bmatrix} 
\bX^H\bX & \bX^H\bv \\
\bv^H\bX & \bv^H\bv
\end{bmatrix}
\end{aligned}
\end{equation}

Using the inverse of a partitioned matrix, the updated inverse is achieved as follows~\cite{khan2008updating}
\begin{equation}
\bY = \begin{bmatrix} 
\bX^H\bX & \bX^H\bv \\
\bv^H\bX & \bv^H\bv
\end{bmatrix}^{-1} = \begin{bmatrix} 
\bF & -\alpha\bY\bX^H\bv \\
-\alpha\bv^H\bX\bY^H & \alpha
\end{bmatrix}
\label{equ:inverse2}
\end{equation}
where

\begin{equation}
\begin{aligned}
\alpha &= \frac{1}{\bv^H\bv-\bv^H\bX\bY\bX^H\bv},\\
\bF &= \bY + \alpha\bY\bX^H\bv\bv^H\bX\bY^H.
\end{aligned}
\label{eq.upd2}
\end{equation}

Equation~\eqref{equ:inverse2} is then used in Algorithm~\ref{SOMP1} to alleviate the computational complexity of taking the full inverse.

\ifCLASSOPTIONcaptionsoff
\newpage
\fi

\bibliographystyle{IEEEtran}
\bibliography{ref}
\end{document}